\documentstyle[preprint,aps]{revtex}
\tightenlines
\begin{document}
\title{Instability of the two-dimensional metallic phase to parallel
magnetic field}

\author{V.M. Pudalov,$^{1,2}$ G. Brunthaler,$^1$ A. Prinz,$^1$
 and G. Bauer$^1$}
\address{$^1$Institut f\"ur Halbleiterphysik, Johannes Kepler
Universit\"at Linz, A-4040 Linz, Austria\\
$^2$Institute for High Pressure Physics, 142092 Troitsk, Russia, e-mail:
pudalov@ns.hppi.troitsk.ru
}
\maketitle
\begin{abstract}
We report on magnetotransport studies of the unusual two-dimensional
metallic phase in  high mobility Si-MOS structures.  We have observed
that the magnetic field applied in the 2D plane suppresses the metallic
state, causing   the resistivity
to increase dramatically by more than 30 times.
Over the total  existence range of the
metallic state, we have found  three distinct types of the
magnetoresistance, related to the corresponding quantum
corrections to the conductivity. Our data suggest that the unusual metallic
state is a consequence of both spin-  and Coulomb-interaction effects.
\end{abstract}

Recently,  convincing evidence for the existence of a  2D metallic state
in Si-MOS structures at zero
magnetic field  has been obtained  in studies of the quantum Hall effect to
insulator transitions \cite{Puda93} and of the Global Phase Diagram
\cite{Krav95}. The
extended states, which in high magnetic field $H$ are centred in the
corresponding Landau bands,   were found  experimentally  to merge and
remain in a finite energy range as $H$ approaches 0, thus providing direct
transitions from the high-order quantum Hall effect states to the insulator
\cite{Puda93}. This  behavior could not  be expected in the framework of the
one-parameter scaling theory  (OPST) \cite{Abra79},
where the extended states are
anticipated to  "float up" in energy as  $H \rightarrow 0$ \cite{Khme84}.
The experimental findings  thus  prove the existence of a mobility edge,
whereas the predicted floating  would evidently correspond to complete
localization.  In subsequent direct studies \cite{Krav94,Krav95b},
the conductivity in
high-mobility Si-MOS structures in zero magnetic field was
found to scale
with temperature and electric field, and  the scaling parameter
demonstrated a pronounced
critical behavior appropriate for a metal-insulator transition.

The  observations of the metal-insulator transition at zero magnetic field
in two dimensional system  raised two major questions: (i) what is the
origin of this unforeseen transition,  and (ii) whether or not  the one
parameter scaling theory  \cite{Abra79} is correct in predicting
the absence of the metallic state in two dimensions. The majority of
experimental data on 2D systems, in fact support the results of
calculations of the "quantum corrections" to the classical diffusion
\cite{Alts85,Fuku85,loc96} rather than the scaling theory in total.

Recently, the strong influence of the  in-plane magnetic field on the
resistivity has been found in Si/SiGe superlattices
\cite{Brun96} as well as in
high mobility Si-MOS structures \cite{Simo97}.
In the current work we report  the
new experimental evidence for the origin of the metal-insulator transition
in Si-structures, and test the applicability of the weak localization
corrections.
We have observed that the magnetic field applied in the 2D plane
destroys the metallic state and restores
the weakly or strongly localized regimes.
Over the existence range of the metallic state,
we have found {\it three distinct types} of the
magnetoresistance  related to the corresponding
quantum corrections  due to interference and interactions
\cite{Alts85,Fuku85}.

The magnetotransport measurements were performed by a 4-terminal
dc-technique.  Four Si-MOS
structures were studied: Si-15A with  peak mobility (at
0.3 K) $\mu  = 41,000$ cm$^2$/Vs,  Si-2Ni with $\mu  = 38,000$, Si-22 with
$\mu  = 26,000$ ,  and Si-39 with $\mu   = 5,000$. While the
first three samples exhibited the metal-insulator
transition \cite{Krav94,Krav95b} and a sharp drop  in resistance
at  $T<2 - 3$K,  the latter low-mobility sample
does not  show  a substantial decrease in resistance.

Fig. 1 shows a set of resistivity curves at different temperatures,
typical for high mobility samples \cite{Krav94}. At carrier density
higher than the critical density, $n_c$ (indicated by an arrow), the
resistance {\em increases} with temperature, while at lower densities it
{\em decreases}. The interception point is slightly dependent on temperature.
The corresponding separatrix between the metallic and insulating sets
of $\rho(T)$ -curves in Fig. 2,a is rising as  $T$ decreases.

Fig. 2,a represents the "metallic" (or high density) part of the
$\rho(T,n_s)$ -plot  and shows a strong drop (by $5 \times$)
in the resistivity below $\sim 2$ K.  As $T$ approaches 0,
$\rho(T)$ saturates and does not show a
tendency to increase, down to at least 14 mK.
The lowest mobility sample Si-39 does not display a
decrease in $\rho$ apart from a few percents in the range 4 to 0.02 K;
the latter behavior  agrees completely with OPST.

{\bf Effect of the magnetic field parallel to the 2D plane.}
The application of an in-plane magnetic field results in a dramatic
increase of the resistance, more than an order of magnitude,
as seen in Fig.2,b. At high fields,  the resistance
saturates. This behavior was found in all three high mobility samples,
in agreement with the results of Ref.\cite{Simo97}.

At high electron densities, the saturation level $\rho^*(H =12.5$ T, $T
\rightarrow 0)$  seems similar to the saturation level at high temperatures
and zero field, $\rho^*(H=0,T = 6$ K), i.e. to the  resistivity anticipated
in the OPST-like behavior. Thus, the magnetic field  simply  destroys
the metallic state.  Comparison of the two
plots, Figs. 2,a and 2,b reveals a remarkable similarity between the effect
of the temperature and of the magnetic field on the resistivity
at high densities.  Both factors destroy the metallic state, and restore
the weakly  or strongly localized regimes.
At densities lower than $2\times 10^{11}$ and closer to
the critical density $n_c$,  the  magnetic
field also gives rise
to an additional $10\times$ larger positive magnetoresistance.

It has been noticed earlier \cite{Puda96} that the temperature
dependence of the
resistivity of the 2D metallic phase may be well described
by an empirical law
$\rho(T) = \rho_1 + \rho_2 exp (-T^{*}/T)$,
where $\rho_1$ is related to scattering at $T=0$,
while the second term  is associated with an energy gap,  $\Delta = kT^{*}$.
Since the parallel field does not affect orbital electron motion, the
magnetic  field may couple to the 2D electrons  only via their spins.
Our results  therefore,  point out
the spin - related origin of the unusual metallic state,
and of the energy gap, $\Delta$.

{\bf Weak localization corrections.}
In the weak perpendicular magnetic field, $H < 0.1$ T, all three high mobility
samples exhibited the weak magnetoresistance, similar to the earlier
reported data \cite{Bish82}.  The
narrow peak in $\rho(H)$  seen in Fig.3 is sensitive
to the normal component of the field and is missing when the field
is aligned with the 2D plane within  $\pm 7$ minutes. Its amplitude does
not vary  much  with the density in the range $(9 - 100) \times 10^{10}$
cm$^{-2}$. These features allow us to attribute the narrow peak to
the orbital single-particle quantum interference correction.  At higher
fields, $H>0.2$ T,  and at high density, $n > 4 \times 10^{11}$ cm$^{-2}$,
the  positive parabolic magnetoresistance dominates, both in parallel and
perpendicular field orientation. This indicates the spin-related origin of
the positive magnetoresistance  component. In the perpendicular field,
the negative  magnetoresistance
grows as the density decreases,  takes over the positive one, and
eventually becomes so large that prevents the observation  of the quantum
interference peak. The negative magnetoresistance persists to the
insulating range of densities, where it was explained by a field effect on
the tunneling conductance \cite{Vois95}.
The negative magnetoresistance is
not seen for a parallel field and is therefore related to the orbital
electron motion.

The positive parabolic magnetoresistance is usually considered as a quantum
correction due to the interaction associated with the Zeeman splitting,
while the negative  magnetoresistance is associated with  a correction due
to electron-electron correlations \cite{Alts85,Fuku85}.  The transition
from the spin - dominant to the Coulomb dominant interaction occurs at the
density $n^{*} \approx 2.8 \times 10^{11}$ for Si-15A and Si-2Ni,
and $n^{*}= 1.7 \times 10^{11}$  for Si-22. These values are noticeably
higher than the critical density at the mobility edge which is
$n_c= 9.0 \times 10^{10}$ for Si-15A and Si-2Ni,
and  $n_c = 10.2 \times10^{10}$ for Si-22.  Therefore,
 the spin-effects and, partly, the Coulomb effects
govern the resistivity over  the existence range of the metallic
phase.

The  persistence  of the quantum corrections to the
conductivity over the total range of existence of the metallic state (see
Fig. 3) seems to justify  the applicability of  the quantum corrections
approach  to the unusual 2D metal.
On the quantitative side,  if we attribute the positive magnetoresistance
(shown in Fig. 2,b)  to the Zeeman-interaction term in the quantum
corrections, than we come up with the conclusion that the
interaction-related quantum corrections are "blowing up" in the vicinity
of the metal-insulator transition
giving rise to the enhancement  factor up to about $10^2 \times$  to the
$\Delta \rho(H)/\rho$ values. This is not surprising since the relevant
theoretical calculations were done in the limit of  $k_F l \gg 1$ where the
corrections are small, whereas in the vicinity of the metal-insulator
transition, at $k_F l \sim 1$,
the quantum corrections may become large.

{\bf Discussion.}
Considering the possible features in which the high mobility Si-MOS
structures differ  from other systems, like GaAs/Al(Ga)As where the
mobility edge was  not found  \cite{Dahm96}, we would like to note
the following:
(i) the Coulomb interaction  energy $E_{ee} = e^2/\kappa r$    is higher in
Si-MOS structures  than in GaAs samples (at  the same interelectron
distance,  $r$) by a factor of 1.7 due to the smaller dielectric constant
 $\kappa = 7.7$ at the Si/SiO$_2$ interface \cite{Puda93},  (ii) the
Si/SiO$_2$ interface is characterized by a very strong asymmetry of the
confining potential in the $z-$ direction. The latter results in a large
effective Lorentzian field $H^{*}$ seen by electrons; the corresponding
spin-orbit gap at zero field was found  to be  equal
to $\approx 4$K \cite{Puda96}.
These effects associated with the broken reflection symmetry of the
confining potential are much less pronounced in GaAs/Al(Ga)As
heterojunctions and are apparently absent in
the rectangular potential wells.

It is known that the inclusion of the spin changes the universality class
of the 2D system. The corresponding scaling consideration is based only on
the symmetry arguments and should not depend much on the particular
microscopic mechanism. The above spin-related mechanism may be
important if the relevant energy gap, $\Delta = g \mu H^{*}$ is larger
than  $\Gamma=h/\tau$, the spin-level broadening.
 It appears, that  the  $\Delta/\Gamma$ -ratio is  $\approx 3$ for
Si-15A and  Si-2Ni,  while   $\Delta/\Gamma
\approx 1$ for the low-mobility sample Si-39, which exhibits normal scaling
behavior and no metal-insulator transition.

Thus, based on the data presented here we suggest that the metallic state and
metal-insulator transition in the high mobility Si-MOS structures
may be  a consequence of both,
the  spin- and  the Coulomb interaction effects.
The former ones are
enhanced  by the broken reflection symmetry of the confining potential
well, while the latter provide the necessary large relaxation time at the
low electron density range.

Recently, some alternative suggestions  on the origin of the unusual 2D
metallic state
in Si-MOS structures were made, namely that it may be induced
by Coulomb interaction \cite{Efro96}, or by spin-triplet pairing
\cite{Beli97}, or that it might be a manifestation of non-Fermi-liquid
behavior \cite{Dobr91}.

One of the  authors (V.M.P.)  benefited from fruitful discussions with
V. Kravtzov and  I. Suslov. The authors acknowledge
support by the Russian Foundation for Basic Research (grant 97-02-17387),
by the Russian State Committee for
Science and Technology (in the framework of  the Programs on the "Physics
of Solid-state Nanostructures" and "Statistical Physics"),
by NWO the Netherlands, and by FWF Austria.

\begin{figure}
\caption[]{Resistivity vs gate voltage
measured on sample Si-22.  Different symbols correspond to 11
temperature values. The electron density is related to the gate voltage
by $n =1.205 \times 10^{11}(V_g - 0.4)$, where $n$ is in cm$^{-2}$ and
$V_g$ in Volts. }
\end{figure}

\begin{figure}
\caption[]{(a) Resistivity vs temperature for the metallic range
of densities, measured on sample Si-15A at zero field.
Different curves correspond to
electron densities between 0.83 and $3.72 \times 10^{11}$ cm$^{-2}$.
(b) Resistivity vs parallel magnetic field, measured at
$T = 0.29$ K on sample Si-15A. Different symbols correspond to
the gate voltages from 1.55 to 2.6 V, or, equivalently, to the densities
from 1.01 to $2.17\times 10^{11}$ cm$^{-2}$.}
\end{figure}

\begin{figure}
\caption[]{Normalized magnetoresistance $(\rho(H)-\rho(0)) / \rho(0)$ vs
perpendicular magnetic field for different densities on
sample Si-2Ni at $T= 1.48$ K. The curves labeled 1 to 12
correspond to the density values of 0.90, 0.96, 1.12, 1.34,
1.56, 2.01, 2.12,
 2.67, 3.77, 4.88, 5.98, and $7.09 \times 10^{11}$ cm$^{-2}$.
The curves were shifted vertically by 0.01 to each other.}
\end{figure}

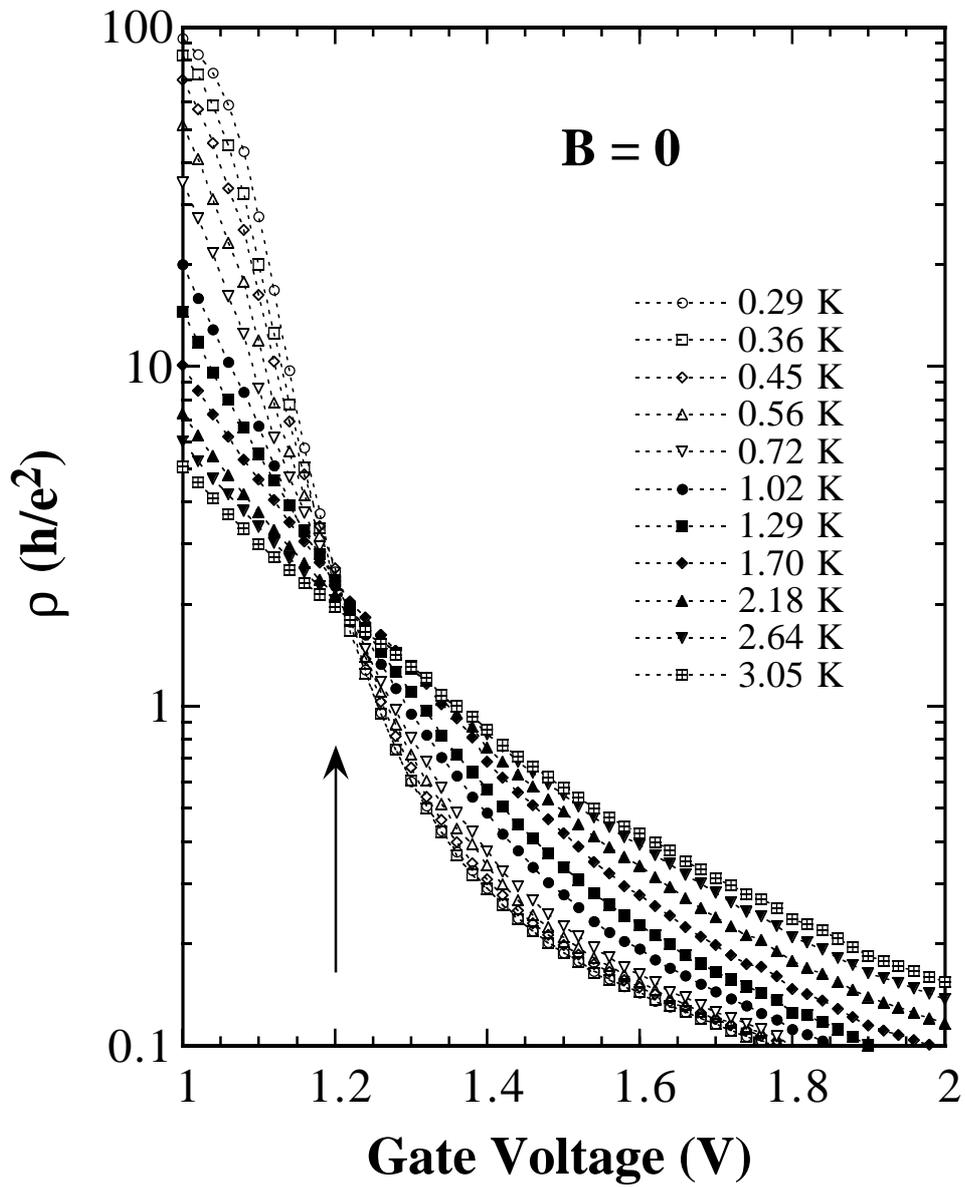
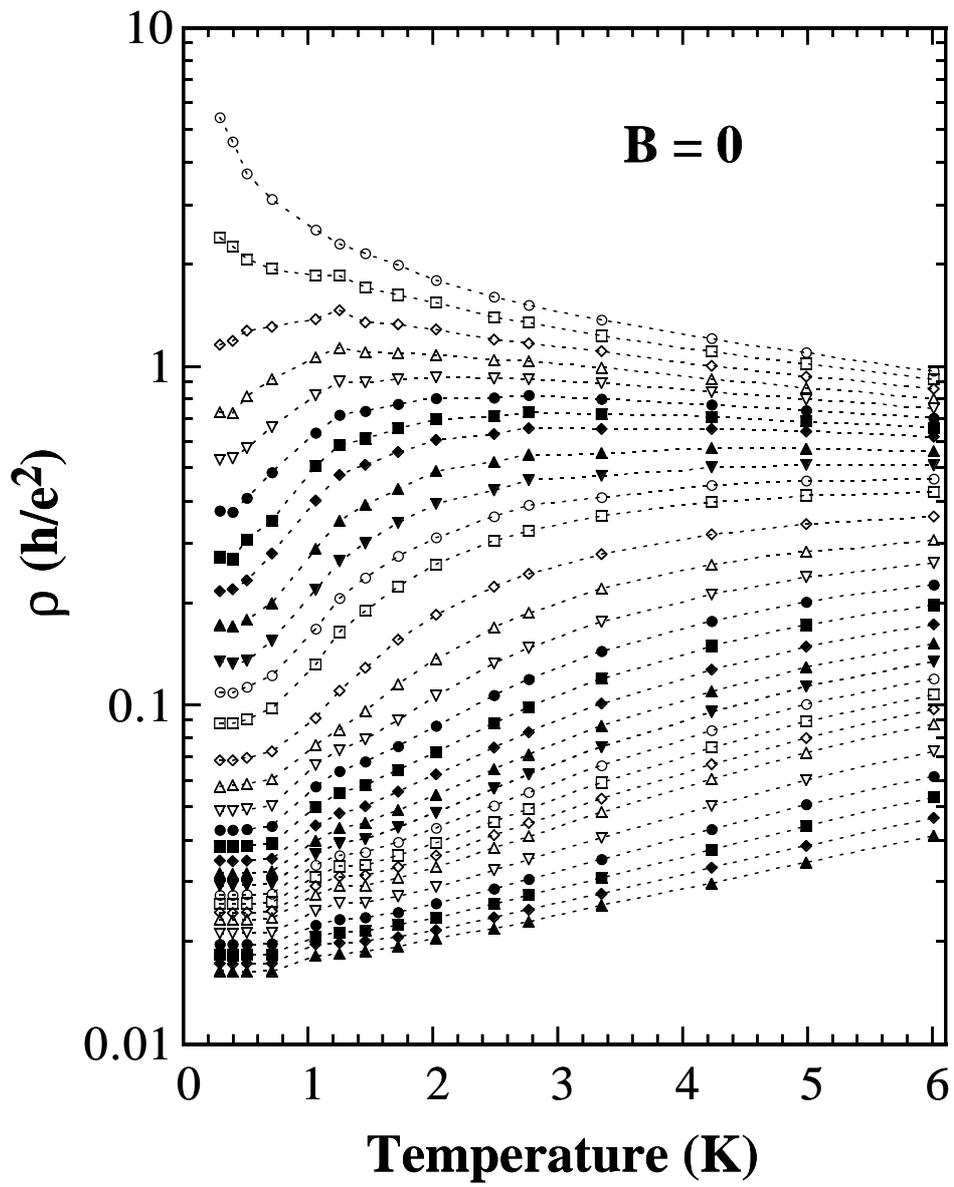
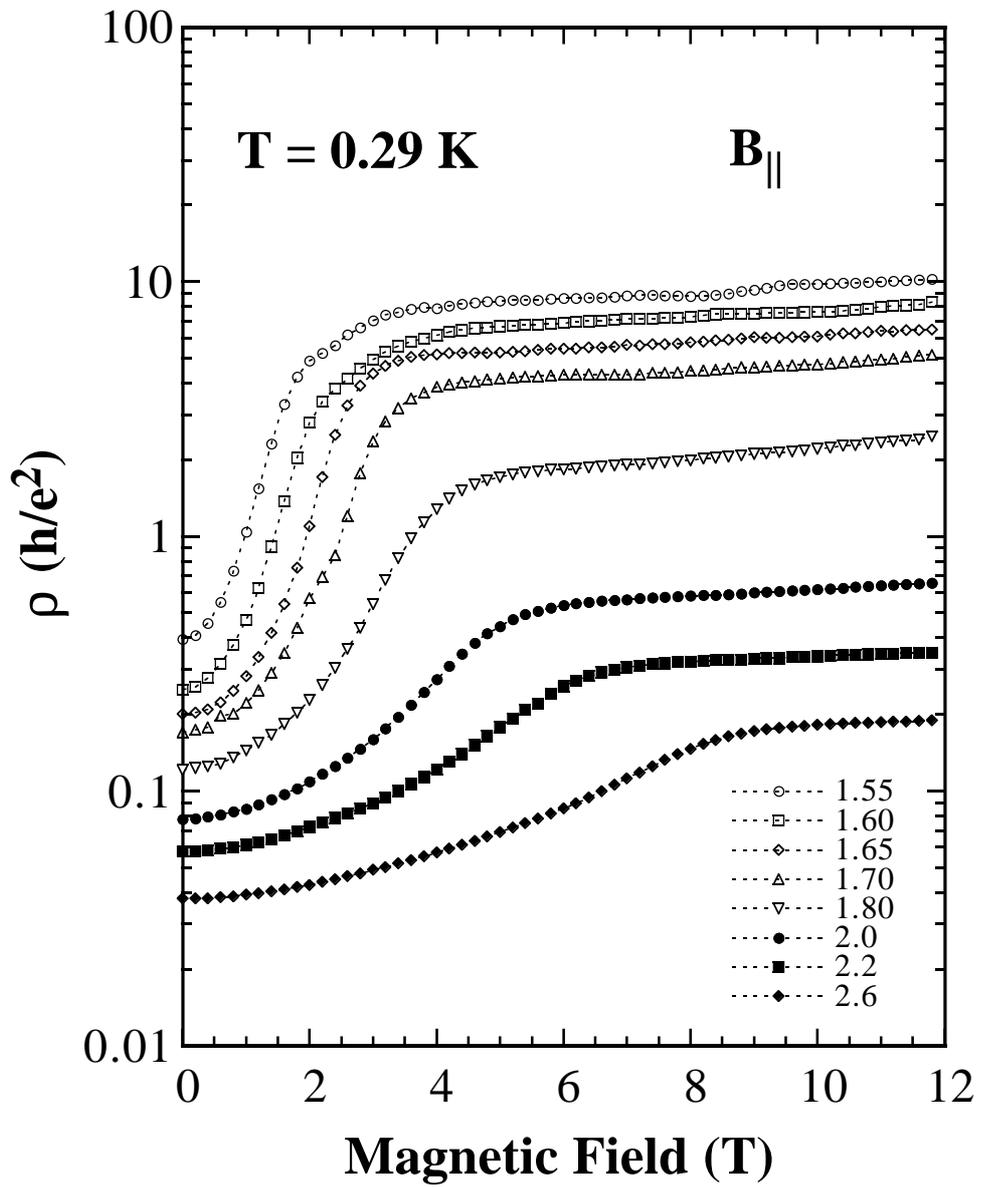
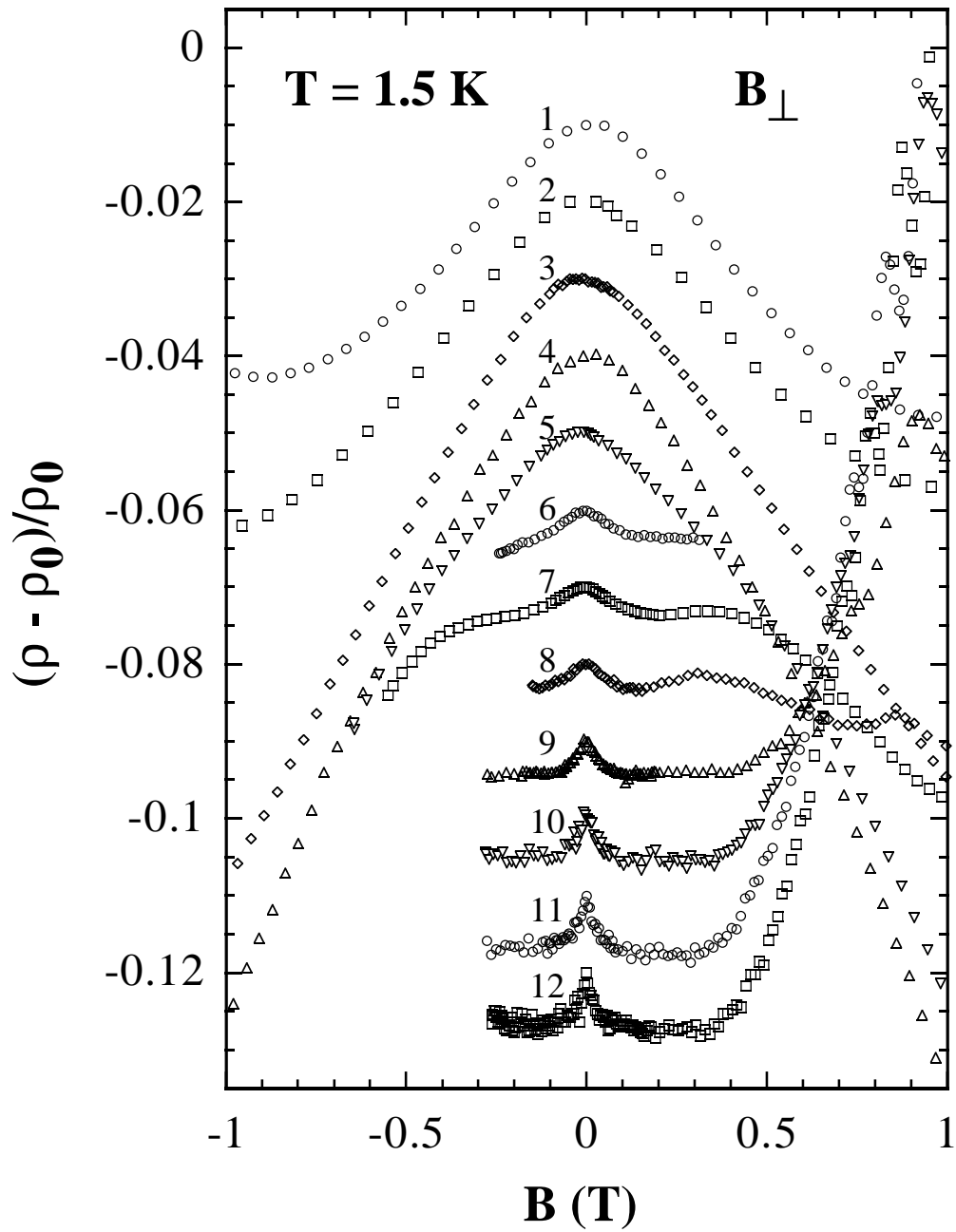
\end{document}